\documentclass[12pt]{iopart}
\usepackage {latexsym}
\usepackage {graphicx}
\usepackage{iopams}
\usepackage {color}

\begin{document}

\title{A Bayesian parameter estimation approach to pulsar time-of-arrival analysis}
\author{C.~Messenger$^1$, A.~Lommen$^2$, P.~Demorest$^3$ and S.~Ransom$^3$}
\address{$^1$ Albert Einstein Institut, 38 Callinstra{\ss}e, Hannover,
30167, Germany}
\address{$^2$ Franklin \& Marshall College, Lancaster, Pennsylvania, USA}
\address{$^3$ National Radio Astronomy Observatory, Charlottesville,
    Virginia, USA}
\ead{chris.messenger@aei.mpg.de}

\begin{abstract}
  The increasing sensitivities of pulsar timing arrays to ultra-low
  frequency (nHz) gravitational waves promises to achieve direct
  gravitational wave detection within the next 5-10 years.  While
  there are many parallel efforts being made in the improvement of
  telescope sensitivity, the detection of stable millisecond
  pulsars and the improvement of the timing software, there are
  reasons to believe that the methods used to accurately determine
  the time-of-arrival (TOA) of pulses from radio pulsars can be improved
  upon.  More specifically, the determination of the uncertainties
  on these TOAs, which strongly affect the ability to detect GWs
  through pulsar timing, may be unreliable.  We propose two Bayesian
  methods for the generation of pulsar TOAs starting from pulsar
  ``search-mode'' data and pre-folded data.  These methods are applied
  to simulated toy-model examples and in this initial work we focus on
  the issue of uncertainties in the folding period.  The final results
  of our analysis are expressed in the form of posterior probability
  distributions on the signal parameters (including the TOA) from a
  single observation.
\end{abstract}

\pacs{04.30.Tv, 95.30.Sf, 95.55.Ym, 97.60.Gb}
\submitto{\CQG}
\maketitle

\section{Introduction}\label{sec:introduction}

Pulsar timing arrays could well be used to detect ultra-low frequency
gravitational waves (GWs) within the next 5-10
years~\cite{2010IAUS..261..228H}.  This is an especially exciting
prospect given the concurrent efforts of the LIGO-Virgo Scientific
collaboration (LVC) whose aim is to make direct detection of GWs (in
the $\sim$10-1000 Hz regime) using the 2$^{\mathrm{nd}}$ generation of
ground based interferometric detectors within the same
timescale~\cite{2009RPPh...72g6901A}.

In this work we outline the beginnings of a Bayesian approach to the
detection of GWs with pulsar timing using simplistic signal and noise
models onto which can be built further levels of sophistication in the
future.  A key long-term aim of our analysis is to improve our ability
to time existing millisecond pulsars by a factor of
3-10~\cite{2009arXiv0909.1058J,2006ChJAS...6b.169H}.  One of the main
problems to be overcome is to be able to sensibly account for the
excess low-frequency noise seen in many stable millisecond
pulsars~\cite{2010CQGra..27h4013H}.  We focus on a \emph{single piece}
of the complete pulsar timing analysis, the generation of
time-of-arrival (TOA) measurements.  Given a single pulsar
observation\footnote{We discuss in Sec.~\ref{sec:discussion} that
while TOAs are associated with individual pulsar observations (or
subsets of an observation), in general a given TOA will depend on
parameters ``fit'' to previous observations.}, this is the
arrival time of the average pulse at the telescope where in this
context ``average'' means the sum of pulses produced by ``folding''
the data with a periodicity equal to the assumed pulse period.  It is
from these TOAs that pulsar astronomers then model the spin evolution
of pulsars taking into account the motion of the radio telescope
relative to the pulsar~\cite{2006MNRAS.369..655H}.  The presence of
GWs in the field between the telescope and the pulsar will result in
small shifts in the arrival times of
pulses~\cite{1979ApJ...234.1100D,1983ApJ...265L..39H}.

We choose to limit our investigation to single pulsar observations
(typically $100-1000$s seconds in duration) and since TOAs are defined in
the reference frame of the telescope and the GW
timescale $\gg$ the timescale of a single observation, we are able to
neglect any GW effect in our analysis.  We will discuss two
different strategies for the estimation of parameters (including the
TOA) from two separate starting points, what we will call
``search-mode'' data and ``pre-folded'' data.  In both cases we
perform the analysis using a commonly used Bayesian integration algorithm
in order to obtain posterior probability distributions on the signal
parameters.

We note that our approach is aimed as a starting point for future more
realistic scenarios and that it can be viewed as an approach being
built from the bottom-up.  We mean this in the sense that we try to
start from the most basic datasets available (see
Secs.~\ref{sec:searchmode} and \ref{sec:folding}) and attempt to build
a data-analysis framework in which the multitude of physical processes
affecting pulsar signals can be included and accounted for.  In
contrast, other work on the specifics of GW detection using pulsar
timing arrays has taken a more top-down approach.  These analyses have started with timing residuals, the result
of a fit to the data assuming non GW effects (effectively the end of
the pulsar data processing chain), and either neglected
this potential
inconsistency~\cite{2004ApJ...606..799J,2009PhRvD..79h4030A} or made
attempts to account for it~\cite{2009MNRAS.395.1005V}.  

The paper is organised as follows.  In Sec.~\ref{sec:signal} we
describe our simplistic signal and noise model.  In
Secs.~\ref{sec:searchmode} and \ref{sec:folding} we then go on to
describe the form of this signal model in two different
representations of the original dataset.  The basic concepts concerning
our Bayesian approach to the parameter estimation problem can be found
in Sec.~\ref{sec:Bayesian} and finally we discuss our conclusions and
potential future developments in Sec.~\ref{sec:discussion}.

\section{The signal : A toy model}\label{sec:signal}
\begin{figure}
  \begin{center}
    \includegraphics[width=\columnwidth]{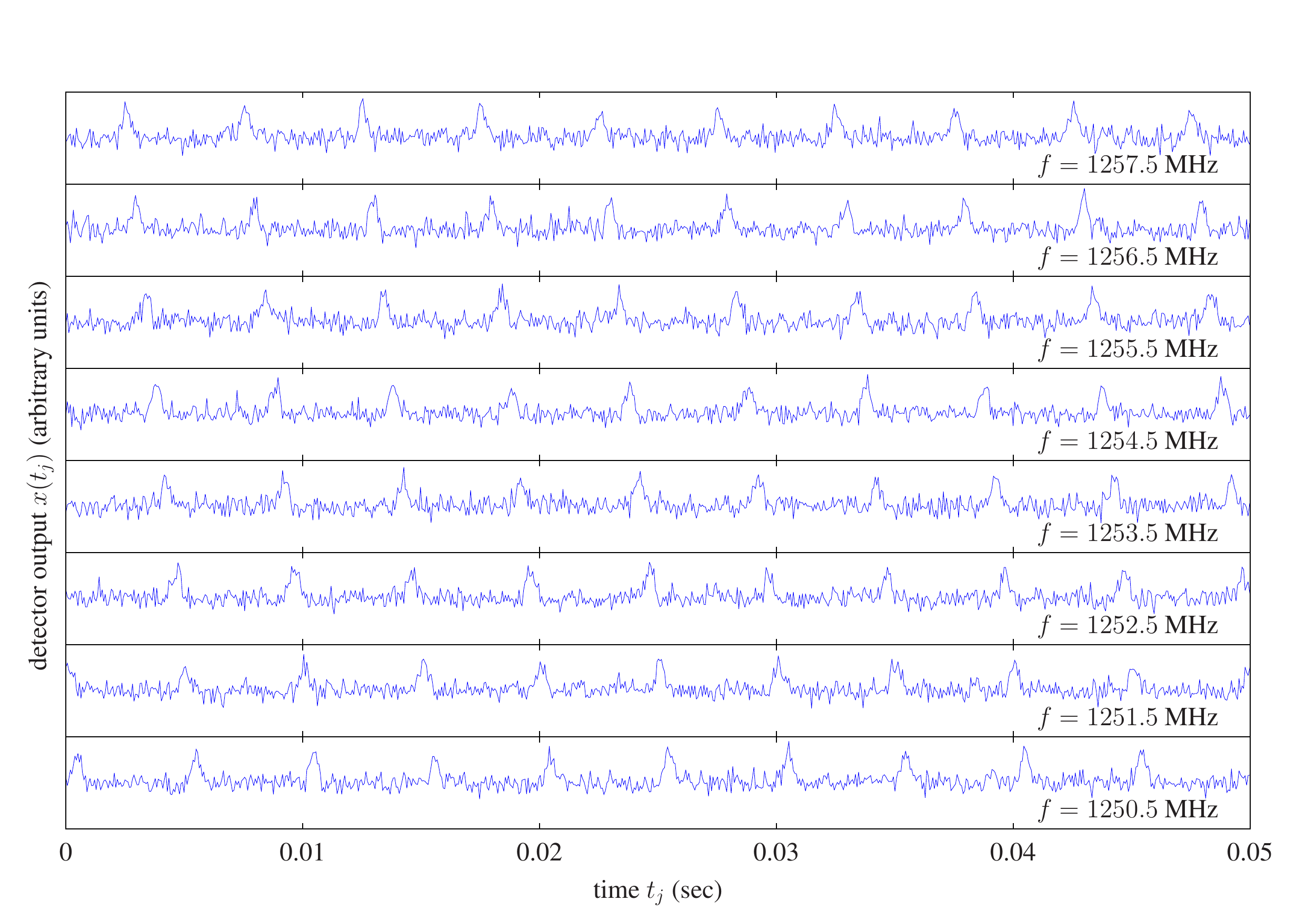}
    \caption{A example of time vs frequency channel ``search-mode''
      data showing a portion of a simulated
      dataset consisting of a strong signal in Gaussian noise.  Here
      we show only the first $0.05$ seconds of data ($\Delta t=64\,\mu$sec) for $8$
      $1$-MHz wide frequency channels.  The signal has an amplitude $A=5$, pulse
      width $w=0.25$ msec, period
      $P=5$ msec, a phase $\Phi_{0}=0.2$, and a dispersion
      measure $DM=100\,\mathrm{cm}^{-3}\mathrm{pc}$.  The noise has unit variance.
    }\label{fig:timedomain}
  \end{center}
\end{figure}

We begin with a dataset defined on a discrete 2-dimensional grid of
time $t_{j}$ versus radio-frequency $f_{k}$ of which an example is
shown in Fig.~\ref{fig:timedomain}.  Data of this kind is often
referred to as
``search-mode'' data since this data format is the kind used when
performing searches for unknown pulsars.  Each of the $M$ rows of the
2-dimensional grid is a timeseries of radio-frequency power measured
within a radio-frequency band with central frequency $f_{k}$.  Typical
sampling times and
observation durations are $\sim 10$s of $\mu$seconds and $100-1000$s
of seconds respectively.  Typical frequency channel widths and total
detector bandwidths are $\sim 1$ MHz and $100-1000$s MHz respectively.
For ``search-mode'' data we assume the following signal model
\begin{equation}\label{eq:signalplusnoise}
  x(t_{j},f_{k}) = s(t_{j},f_{k}) + n(t_{j},f_{k}),
\end{equation}
where $x(t_{j},f_{k})$ represents the discretely sampled dataset,
$s(t_{j},f_{k})$ is the signal and $n(t_{j},f_{k})$ is the noise which
for simplicity we assume as independent Gaussian distributed random
variables with zero mean.   The signal itself we
define as
\begin{equation}\label{eq:signalmodel}
  s(t_j,f_k) =
  \sum_{\alpha=0}^{n'-1}A\exp\left[-\frac{\left(t_{j}-\mu_{\alpha
        k}\right)^2}{2w^2}\right],
\end{equation}
where $\alpha$ sums over all $n'$ pulses that intersect with the
observation\footnote{Due to the effects of dispersion, whilst the
  pulse period is equal in all frequency channels, a particular pulse
  near the end of the timeseries for a high-frequency channel can be
  delayed by dispersion such that it does not intersect with the
  observation at a lower frequency channel.  The same applies to
  pulses near to the start of the timeseries in a lower frequency
  channel since they may arrive before the observation in a higher
  frequency channel.}.  We use $A$ as the pulse peak amplitude, $w$ as
the pulse width, and $\mu_{\alpha k}$ as the centre of the $\alpha$'th
pulse in the $k$'th frequency channel.  Note that we are modelling
each pulse as having a single Gaussian profile component and that the
amplitude and width remain constant in both time and with frequency
channel.  The inclusion of additional Gaussian pulse components
requires only a
trivial modification to the model.  In Sec.~\ref{sec:discussion} we discuss numerous potential
additions and modifications required to make this toy model a more
accurate representation of a real pulsar signal.

The time at the centre of each pulse is defined as
\begin{equation}\label{eq:meanpulse}
  \mu_{\alpha k} = \left(\phi_{k}+\alpha\right)P + \xi_{\alpha},
\end{equation}
where $P$ is the constant pulse period and $\phi_{k}$ is the phase
(defined on the range $\left[0,1\right)$) of the first pulse in the
observation for the $k$'th frequency channel.  We have also included a
random pulse ``jitter'' term where for each pulse we apply a random
shift to the pulse arrival time where each shift $\xi_{\alpha}$ is
drawn from a Gaussian distribution with zero mean and variance
$\sigma_{\xi}^{2}$.  Such effects have been observed in several
pulsars and can be attributed to unknown processes in the pulse
emission mechanism and possibly related to giant
pulses~\cite{2000ApJ...535..365K,2002MNRAS.334..523K,1996ApJ...457L..81C,2007A&AT...26..585K,2006AstL...32..583K}.
We show in Sec.~\ref{sec:searchmode} that for our purposes, the effect
of this particular pulse ``jitter'' modelling can be absorbed into a
subset of the other signal parameters.
  
  The phase of the first pulse in each frequency channel $\phi_{k}$
  can be related to the phase $\Phi_{0}$, defined as the phase of the
  pulse at the midpoint frequency channel
  $f_{\mathrm{mid}}=(f_{M}-f_{1})/2$ and with reference to the
  midpoint of the observation $t = T/2$ by %
  \begin{equation}\label{eq:phi}
    \phi_{k} = \mathrm{mod}\left(\frac{T}{2P} + \frac{\Delta
      t(f_{k})}{P} + \Phi_{0},1\right).
\end{equation}
The relative delay due to dispersion in the $k$'th frequency channel
$\Delta t(f_{k})$ is given by
\begin{equation}\label{eq:deltat}
  \Delta t(f_{k}) = 4.148808\times 10^3
  \left(f_k^{-2}-f_{\mathrm{mid}}^{-2}\right) DM\,\,\,\mathrm{sec},
\end{equation}
where $DM$ is the dispersion measure in $\mathrm{cm}^{-3}\mathrm{pc}$
and the units of the frequencies are MHz.  Note that in our simplistic
model we do not account for dispersion smearing within individual channels.
%
\section{Using ``search-mode'' data in the Fourier domain}\label{sec:searchmode}

\begin{figure}
  \begin{center}
    \includegraphics[width=\columnwidth]{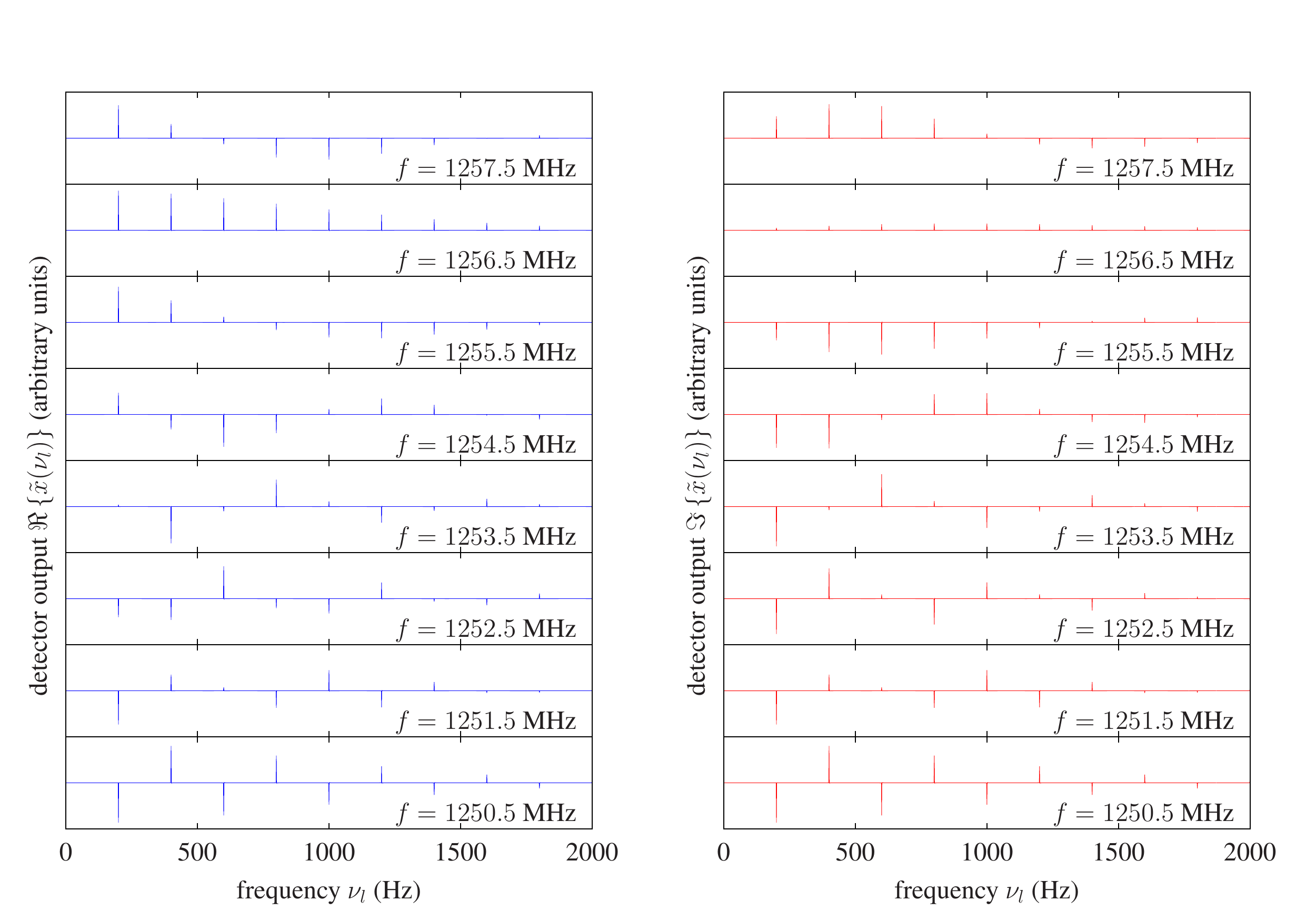}
    \caption{A example of Fourier transformed ``search-mode'' data showing a portion of a simulated
      dataset consisting of a strong signal in Gaussian noise.  The
      panels on the left show (in blue) the real part of the complex Fourier
      transform of the data as a function of Fourier frequency for $8$
      $1$-MHz wide
      frequency channels.  The imaginary parts are shown (in red) on the
      right.  The dataset used to generate this plot is identical to that shown in
      Fig.~\ref{fig:timedomain} and we have truncated the frequency
      range at $2$ kHz since there the harmonic content of the signal is
      significantly reduced beyond this frequency.
    }\label{fig:frequencydomain}
  \end{center}
\end{figure}

The signals received from pulsars are periodic and their frequency
evolution is slow i.e. the timescale of frequency variation $\gg$ the
pulse period.  By Fourier transforming each channel's time-series
we find that a pulsar signal can be represented as a series of
narrow-band harmonics as shown in Fig.~\ref{fig:frequencydomain}.   In
a realistic situation we would expect to have some prior
knowledge of the pulsar frequency before performing an analysis and
therefore transforming the data
in such a way allows us to isolate the regions in the dataset where
the signal is concentrated (at the harmonics).   This in-turn allows
us to be economic with the data samples that we are interested in and will make any numerical
likelihood computation more efficient.
Let us define the discrete Fourier transform as
\begin{equation}\label{eq:fouriertransform}
  \tilde{x}(\nu_{l})=\sum_{j=0}^{N-1}x(t_{j})e^{-2\pi jl/N}\Delta t,
\end{equation}
where $\nu_{l}$ represents the elements of a vector containing the positive
discrete Fourier frequencies\footnote{Note that there is a clear
  distinction between the radio frequencies (or frequency channels)
  and the Fourier frequencies.} with frequency spacing $1/T$
and where $N$ is the number of time samples.
When applied to the time series from each frequency channel
of a noise-free signal (defined by Eqs.~\ref{eq:signalmodel},\ref{eq:meanpulse})
we obtain 
\begin{eqnarray}\label{eq:fourierpulse1}
  \tilde{s}(\nu_{l},f_{k}) &=& \sum_{j=0}^{N-1}\sum_{\alpha=0}^{n'-1}A\exp\left[-\frac{\left(t_{j}-\mu_{\alpha
      k}\right)^2}{2w^2}\right]e^{-2\pi jl/N}\Delta t, \nonumber \\
  &=& \sum_{\alpha=0}^{n'-1}\tilde{s}_{\alpha}(\nu_{l},f_{k}),
\end{eqnarray}
where we have decomposed the complete Fourier transform into the Fourier
transform of each pulse.  We then have
\begin{eqnarray}
  \tilde{s}_{\alpha}(\nu_{l},f_{k}) &=& A\sum_{j=0}^{N-1}\exp\left[-\frac{\left(t_{j}-\mu_{\alpha
      k}\right)^2}{2w^2} - 2\pi jl/N\right]\Delta t, \nonumber \\
  &\approx& A\exp\left\{-2\pi i \nu_{l}\mu_{\alpha k}\right\}\int_{-\infty}^{\infty}\exp\left\{\frac{y^{2}}{2w^{2}}-2\pi i
  \nu_{l}y\right\}\,dy,\nonumber \\ 
  &=& Aw\sqrt{2\pi}\exp\left\{-2\pi i\nu_{l}\mu_{\alpha
      k}-2\pi^{2}w^{2}\nu_{l}^{2}\right\},\label{eq:fourierpulse2}
\end{eqnarray}
where we have approximated the discrete sum over time samples with the
continuous integral over the dummy variable $y=t-\mu_{\alpha k}$
assuming that each pulse itself spans $\gg 1$ time bin and is not
truncated by the edges of the timeseries\footnote{Clearly, of the $n'$
  pulses that intersect the time-frequency plane there will
  be some frequency channels for which a pulse does not appear in the
  timeseries due to dispersion.  Equation~\ref{eq:fourierpulse2} is
  therefore only applicable to those pulses in a particular frequency
  channel that are found to intersect the timeseries.}.  We can now perform the sum
over $\alpha$ (the individual pulses) to obtain the complete Fourier
transform.  However, note that $\mu_{\alpha k}$ is a function of
$\xi_{\alpha}$, the random individual pulse arrival time
jitter.  We choose to average over this random variable under the assumption that
there are a large number of pulses within the observation
time.  This averaging procedure leads to the following replacement:
\begin{eqnarray}\label{eq:averagejitter}
  e^{-2\pi i\nu_{l}\xi} \rightarrow\Large\langle e^{-2\pi
    i\nu_{l}\xi}\Large\rangle &=& \int_{-\infty}^{\infty} \frac{e^{-\xi^{2}/2\sigma_{\xi}^{2}}}{\sqrt{2\pi\sigma^{2}_{\xi}}}\,e^{-2\pi
  i\nu_{l}\xi}\, d\xi, \nonumber \\
  &=& e^{-2\pi^{2}\nu_{l}^{2}\sigma_{\xi}^{2}},
\end{eqnarray}
where we have replaced the pulse jitter term with its expectation
value and used a Gaussian distribution for the pulse jitter with a
zero mean and variance of $\sigma^{2}_{\xi}$.  Finally we obtain the
following expression for the Fourier transform of the signal only
timeseries
\begin{equation}\label{eq:completefourier2}
  \tilde{s}(\nu_{l},f_{k}) 
  = \frac{A_{\xi}w_{\xi}T}{P}\sqrt{2\pi}\exp\left\{-2\pi^{2}\nu_{l}^2 w_{\xi}^{2}
  \right\}\exp\left\{-2\pi i \phi_{k}\nu_{l}\right\}\tilde{W}_{l}.
\end{equation}
We can see from this equation that in the Fourier domain the signal
can be decomposed into four parts.  There is a real positive amplitude term proportional to
the pulse amplitude, width, and number of pulses ($n\approx T/P$)
which is multiplied by a frequency dependent envelope function that
decays with increasing frequency at a rate proportional to the pulse width.  There is also a unit amplitude complex phase term
dependent upon the initial phase of the pulse in the given frequency
channel multiplied by a second complex phase term
$\tilde{W}_{l}$ given by 
\begin{equation}
  \tilde{W}_{l} = \frac{P}{T}\exp\left\{2\pi i \nu_{l}P\right\}\left[\frac{1 -
      \exp\left\{-2\pi i\nu_{l}T\right\}}{\exp\left\{2\pi
      i\nu_{l}P\right\} - 1}\right],
\end{equation}
which, in the limit of $T\gg P$ can be written as 
\begin{equation}
  \tilde{W}_{l} = \sum_{\beta=0}^{n}\left\{\frac{\sin(2\pi\Delta\nu_{l\beta}T)}{2\pi\Delta\nu_{l\beta}T} +
  i\left[\frac{\cos(2\pi\Delta\nu_{l\beta}T) -1}{2\pi\Delta\nu_{l\beta}T}\right]\right\},
\end{equation}
where $\Delta\nu_{l\beta}=\nu_{l} - \beta/P$ and $\beta$ labels
the individual signal harmonics of which there are $n$.  This final complex
phase term contains the information regarding the location and phase
of the signal harmonics.  We can now see that each signal harmonic is
identical in shape but will each have a different phase and
amplitude.  In addition, as one moves to different frequency channels
the phase of a given harmonic will be rotated by a quantity dependent upon the dispersion measure. 

Note that we have also re-parameterised the pulse amplitude and width
using 
\begin{eqnarray}
  A_{\xi} &=& \frac{Aw}{\sqrt{w^{2}+\sigma_{\xi}^{2}}}, \label{eq:jitteramplitude}\\
  w_{\xi} &=& \sqrt{w^{2}+\sigma_{\xi}^{2}}, \label{eq:jitterwidth}
\end{eqnarray}
since with the addition of pulse jitter there exists a degeneracy
between the original pulse amplitude and width.  The product of the
amplitude and width determine the overall amplitude of the Fourier
transform of the signal and the sum of the squares of the width and
the pulse jitter parameter determine the rate of the reduction in
amplitude of the harmonics with increasing frequency.  Using the data
to measure this amplitude and its attenuation with increasing
frequency will therefore not allow us to constrain all three
parameters\footnote{We note that strictly speaking it would be
  possible to identify the values of all three parameters for a very
  strong signal.  Pulse arrival time jitter acts to remove a small
  fraction of power from
  the harmonics and distribute it amongst the inter-harmonic frequency
  bins.  Our analysis is restricted to localised regions at each
  harmonic and so we treat this information as lost.}.
%
\section{Using folded data}\label{sec:folding}

The majority of pulsar timing data is pre-processed and reduced in
volume by the
process of folding.  In this process, sections of the time series from each
frequency channel of an observation will be folded with an assumed
pulse period\footnote{The folding procedure can also include
  de-dispersion over a limited range of frequency channels where, just
as with folding, an assumed value of the dispersion is used.  Hence a
large number of frequency channels can be grouped together into a
single pulse profile measurement.  We do not consider this potential feature of
the folding procedure in this work.}.  At the time of folding this pulse period will not
necessarily be the most accurate value.  The pulse period itself is
updated and refined with each subsequent observation.  However, once
data have been folded, most notably for older observations, the
original search mode data may be lost, meaning that re-folding with the more
refined period is not possible.  

We will focus on the effect of
folding with an inaccurate pulse period.  One can argue that since the most basic initial
pulse period estimates will require a coherent measurement over some
prior observation spanning many pulses, we should expect an initial
worst case fractional uncertainty in the pulse period of $\sim P/T$
which for a $10$ msec pulsar period and a $100$ second coherent observation
equates to a period error of $\sim 1$ $\mu$second.  In addition to the pulse
period, for realistic analyses a number of other parameters are used
in the folding procedure such as the sky position coordinates, the
intrinsic pulsar spin-frequency derivatives, the dispersion measure
plus orbital parameters if the source is in a binary system.  In our
toy model we ignore these complications.  

\begin{figure}
  \begin{center}
    \includegraphics[width=\columnwidth]{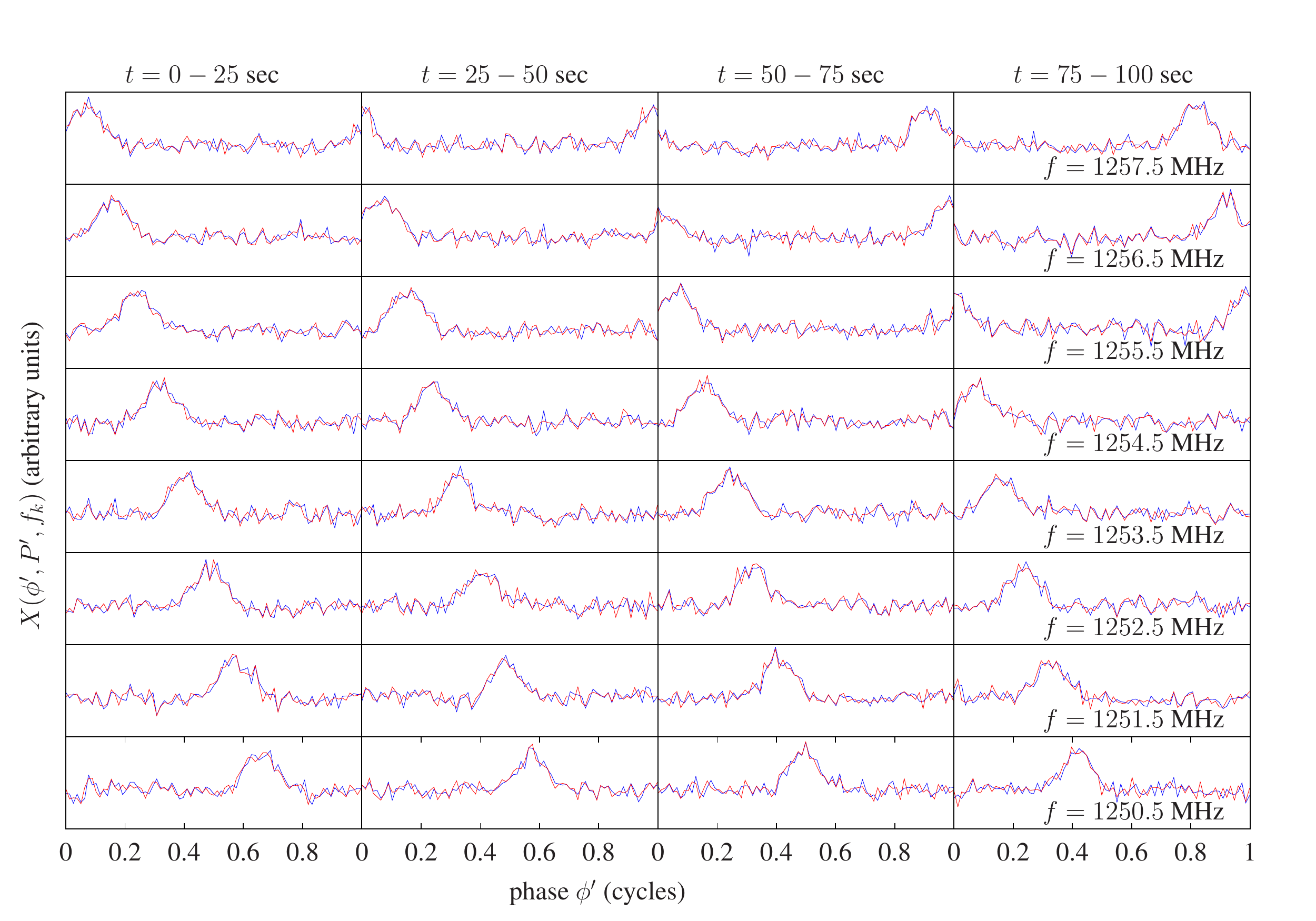}
    \caption{An example simulated folded dataset showing folded pulse profiles
      for $4$ sub-intervals each spanning $25$ seconds and for $8$
      frequency channels each spanning $1$ MHz.  The simulated signal parameters
      are equal to those defined and used in
      Figs.~\ref{fig:timedomain} and~\ref{fig:frequencydomain} with
      the exception that here the signal amplitude $A=0.1$ is significantly
      lower.  Two curves are plotted in each panel, the blue curves
      are profiles obtained through folding with the true pulse period
      $P$.  The red curves are the profiles obtained through folding
      with an pulse period error $\Delta P'=10$ nsec.  Note
      that this size period error is equivalent to a phase error of
      $\sim 0.01$ cycles over the course of a sub-integration.
    }\label{fig:foldeddata}
  \end{center}
\end{figure}

We choose to define the result of the folding process for a single observation as a
2-dimensional grid of pulse profiles labelled by time and channel
frequency, an example of which is shown in Fig.~\ref{fig:foldeddata}.  To perform a consistent analysis of such a dataset we 
take into account the fact that profiles have been obtained using a
non-precise value of the pulse period.  If we consider a dataset that
has already been folded at a specific (non-exact) pulse period
$P'=P+\Delta P$ then we can define a new folded dataset as
\begin{equation}\label{eq:foldeddata}
  X(\phi',P',f_{k}) = \sum_{\beta=0}^{n-1}x\Big((\beta+\phi')P',f_{k}\Big),
\end{equation}
where $\beta$ indexes each fold up to $n=\mathrm{floor}(T/P')$.
Substituting in our signal model (Eqs.~\ref{eq:signalmodel} and \ref{eq:meanpulse}) we
can accurately approximate the discretely summed noise-free pulse profile as
\begin{eqnarray}\label{eq:profilemodel}
 S(\phi',P',f_{k})
 &\approx&
  \frac{A_{\xi}w_{\xi}}{|\Delta P|}\sqrt{\frac{\pi}{2}}\sum_{z=-1}^{1}\left[\mathrm{erf}\left(a_{z}+b\right)
      - \mathrm{erf}\left(a_{z}\right)\right],
\end{eqnarray}
where we have used 
\begin{eqnarray}
  a_{z} &=& \frac{|\Delta P|}{\Delta P}\,\left[\frac{(P+\Delta P)(\phi'+z)-\phi_{k}
      P}{w_{\xi}\sqrt{2}}\right], \label{eq:foldingaz}\\
  b &=& \frac{|\Delta P|(n-1)}{w_{\xi}\sqrt{2}}. \label{eq:foldingb}
\end{eqnarray}
In the calculation of Eq.~\ref{eq:profilemodel} we have again replaced
the pulse arrival time jitter term with its expectation value (as done
in Sec.~\ref{sec:searchmode}) and approximated the sum over pulses
with a continuous integral.  We have also been forced to
re-parameterise the pulse amplitude and width parameters for the same
reasons as described in the previous section and have chosen to use an
identical re-parameterisation (defined in
Eqs.~\ref{eq:jitteramplitude} and~\ref{eq:jitterwidth}).  The summation
over the index $z$ is simply to account for the fact that folding a
signal with an arbitrary initial phase may separate the pulse profile
into significant contributions spanning the $\phi'=0=1$ point.  This
also acts to account for the fact that if folding with an incorrect
pulse period the true pulse will slowly drift across the $\phi'$
space.  In this scenario the tails of neighbouring pulses begin to
contribute to the sum and by including the $z=\pm 1$ terms we are
accurately modelling this effect.
%
\section{A Bayesian analysis}\label{sec:Bayesian}

The Bayesian component to our approach can be viewed as standard in
the sense that we aim to simply apply Bayes probability theorem to the
time-of-arrival problem with the intention of computing marginalised
posterior probability distributions on the signal parameters.  

Bayes theorem can be expressed as
\begin{equation}\label{eq:bayestheorem}
  p(\boldsymbol{\theta}|\boldsymbol{x},\mathcal{M}) = \frac{L(\boldsymbol{x}|\boldsymbol{\theta},\mathcal{M})\pi(\boldsymbol{\theta}|\mathcal{M})}{E(\mathcal{M}|\boldsymbol{x})},
\end{equation}
where the term on the left-hand-side is the joint posterior
probability distribution on the parameter set $\boldsymbol{\theta}$
given a dataset defined by the vector $\boldsymbol{x}$ and a chosen model
represented by $\mathcal{M}$.
The function $L(\boldsymbol{x}|\boldsymbol{\theta},\mathcal{M})$ is the
likelihood function describing the dataset
$\boldsymbol{x}$ given the parameter set $\boldsymbol{\theta}$ and the
model $\mathcal{M}$.  The function $\pi(\boldsymbol{\theta}|\mathcal{M})$ is the joint prior
probability distribution on the parameter set $\boldsymbol{\theta}$
given the model $\mathcal{M}$.  Finally we have the Bayesian evidence
$E(\mathcal{M}|\boldsymbol{x},\boldsymbol{\theta})$ representing the probability
of the model $\mathcal{M}$ given the dataset $\boldsymbol{x}$.

To obtain marginalised posterior distributions on a particular signal
parameter we are required to perform a multi-dimensional integration
of the joint posterior distribution over the remaining parameters.
Formally this can be written as
\begin{equation}\label{eq:posterior}
  p(\theta_{m}|\boldsymbol{x},\mathcal{M}) \propto \int_{\mathcal{S}} d^{n}\boldsymbol{\theta'}\,
  L(\boldsymbol{x}|\boldsymbol{\theta},\mathcal{M})\pi(\boldsymbol{\theta}|\mathcal{M}),
\end{equation}
where the parameter vector $\boldsymbol{\theta'}$ consists of the
subset of parameters in the vector $\boldsymbol{\theta}$ excluding the
parameter $\theta_{m}$ and where $\mathcal{S}$
defines the volume of integration on that space.  Note that there is no
dependence upon the Bayesian evidence in the calculation of the
marginalised posterior distribution since it is independent of the
parameter values themselves and can be absorbed into the normalisation
of the posterior distribution.

In practice the calculation of posterior distribution functions can be
a difficult and computationally intensive procedure.  Over the last
decade much work has been dedicated to the efficient numerical
computation of posterior probability distributions and more recently
to the evaluation of the Bayesian evidence.  One of the now standard
tools available for Bayesian data analysis is the
Markov-Chain-Monte-Carlo (MCMC)~\cite{MCMCinPractice,Gelman95}, an efficient
method for obtaining random samples drawn from a posterior probability
distribution of which there are a number of
variations~\cite{marinari-1992-19,
  Gramacy:2010:IT:1713542.1713556, Cai:2008:MAA:1484982.1485000,
  RePEc:mtn:ancoec:2001:3:16,Hernandez-Marin_2007, 2009arXiv0904.2207T}. More
recently the strategy known as ``nested sampling''~\cite{skilling:395,Sivia96}
has given the data analyst the ability to accurately estimate the
Bayesian evidence, a model dependent quantity used to perform model
selection.  The first direct application of this
strategy was to perform cosmological model
selection using {\it WMAP}
data~\cite{2006ApJ...638L..51M}.  For this work we chose to perform our analysis using the freely
available nested sampling algorithm
\verb1MultiNest1~\cite{2008MNRAS.384..449F}.  Note that this algorithm
has been specifically designed to be robust with respect to
multi-modal posterior distributions and to compute the Bayesian
evidence.  For this work we use it purely to obtain posterior
probability distributions on the pulsar parameters.  

Let us now define the likelihood functions specific to the two
approaches described in Secs.~\ref{sec:searchmode} and~\ref{sec:folding}.
The likelihood function for the Fourier domain approach to the
``search-mode'' data is defined as 
\begin{equation}\label{eq:searchmodelikelihood}
  L^{\mathrm{sm}}(\boldsymbol{\tilde{x}}|\boldsymbol{\theta}) =
  \left(2\pi\sigma_{f}^{2}\right)^{-NM/4}\exp\left\{-\frac{1}{2\sigma^{2}_{f}}\sum_{j=0}^{N/2-1}\sum_{k=0}^{M-1}
   |\tilde{x}_{jk} - \tilde{s}_{jk}(\boldsymbol{\theta})|^2\right\},
\end{equation}
where $N/2$ and $M$ are the total number of Fourier-frequency and radio-frequency
bins respectively and we define $\boldsymbol\theta=\{A_{\xi},w_{\xi},DM,P,\Phi_{0}\}$ as
the vector of signal parameters.  We have used $\sigma_{f}^{2}$ to
represent the frequency domain noise variance which we assume to be
Gaussian, white, and stationary and therefore constant for all Fourier
and radio frequency bins.  In this ideal scenario the frequency domain
noise variance is related to the time domain noise variance
$\sigma_{t}^{2}$ by $\sigma_{f}^{2} = N(\Delta t)^2\sigma_{t}^{2}$.    

The likelihood function for the folded data can similarly be written as
\begin{equation}\label{eq:foldedlikelihood}
L^{\mathrm{fold}}(\boldsymbol{X}|\boldsymbol{\theta}) =
  \left(2\pi\sigma_{X}^{2}\right)^{-N_{\mathrm{s}}M/2}\exp\left\{-\frac{1}{2\sigma^{2}_{X}}\sum_{j=0}^{N_{\mathrm{s}}-1}\sum_{k=0}^{M-1}
   \left(X_{jk} - S_{jk}(\boldsymbol{\theta})\right)^2\right\},
\end{equation}
where $N_{s}$ is the number of equal length sub-intervals into which each frequency
channel's timeseries has been divided.  The noise contribution in
a particular folded phase bin is simply the sum of $n=\mathrm{floor}(T/P')$ Gaussian
distributed variables of variance $\sigma_{t}^{2}$ and hence
$\sigma_{X}^{2}=n\sigma_{t}^{2}$.  The parameter vector
$\boldsymbol{\theta}$ is identical to that defined for the search-mode
data.

In general the choice of prior probability distribution functions on the parameters
$\boldsymbol{\theta}$ would be chosen according to one's prior beliefs
on the values of those parameters.  However, for the purposes of our
toy model investigation we choose ``flat'' prior distributions for all
parameters with prior ranges chosen to be far greater than the
expected span of the posterior distributions.  In this case we do not
favour any particular choice of parameter values over any others.We note that in making
this choice we are disregarding a powerful feature of the Bayesian
analysis, the ability to correctly incorporate prior information into
the result.  However, one can show that for strong signal-to-noise
ratios the effect of the prior on the posterior is dominated by that of the likelihood
function itself. 

To conclude this section we would like to make it clear that what we
have described in Secs.~\ref{sec:searchmode}
and~\ref{sec:folding} do \emph{not} constitute two separate models.
We have described two separate representations of the same original
dataset and have in-fact used the same signal model.  Model selection
therefore could \emph{not} be applied to these two methods.  Our aim
is to compare the effectiveness of each choice of dataset
representation by contrasting the posterior distributions on the
signal parameters when a single common time-radio-frequency dataset is used
to generate both the Fourier-radio-frequency and a folded dataset.
Model selection using the Bayesian evidence and the computation of the
Bayes factor (the ratio of model evidences) and odds-ratio (the Bayes
factor multiplied by the ratio of prior model probabilities) is a
potentially powerful tool in future advanced implementations of our
analysis strategy.  Our choice of nested sampling implementation, \verb1MultiNest1, has
been designed specifically to compute the Bayesian evidence, making
model selection between different pulsar signal models an obvious and
easy to implement extension of our approach.
%
\section{Discussion}\label{sec:discussion}

\begin{figure}
  \begin{center}
    \includegraphics[width=\columnwidth]{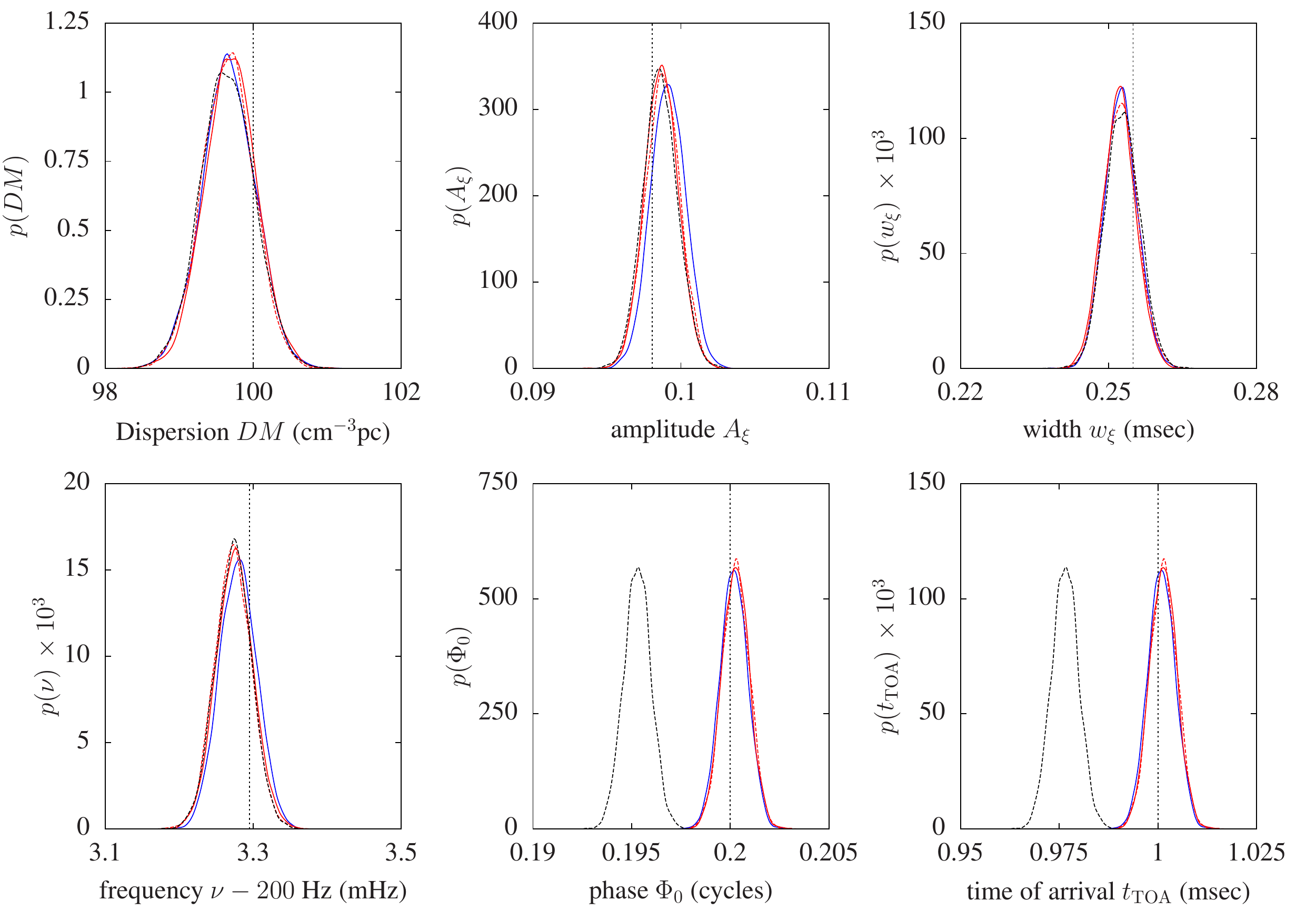}
    \caption{The marginalised posterior distributions on the signal
      parameters for a simulated signal in Gaussian noise.  In solid blue we
      show the results obtained when using the Fourier domain
      representation of the ``search-mode'' data.  In solid red we
      show the results obtained when using folded data as the input
      for the case where the data was folded with the true pulse
      period.  In dashed red we show the results for the folding
      scenario where an incorrect period, $\Delta P=10$ nanoseconds,
      has been used to fold the data and we have accounted for this
      within the signal model.  The dashed black curves show the
      result where this effect has not been accounted for.  The
      vertical dotted black lines indicate the values of the true
      signal parameters.  For all
      results the data was converted into the Fourier and folded
      representations from a single common time-frequency dataset of
      length $100$ sec with sampling time $64$ $\mu$sec and frequency
      range of $8$ MHz consisting of $8$ channels each of $1$ MHz bandwidth.    
    }\label{fig:posteriors}
  \end{center}
\end{figure}

Shown in Fig.~\ref{fig:posteriors} is an example of typical
marginalised posterior probability distributions on the signal
parameters $\boldsymbol{\theta}=\{A_{\xi},w_{\xi},DM,P,\Phi_{0}\}$ plus
the time-of-arrival parameter $t_{\mathrm{TOA}}$.  The latter is not independent of the
other parameters and is a function of both the phase parameter and the
pulse period such that $t_{\mathrm{TOA}} = \Phi_{0}P$ and is therefore
defined as the arrival time of the first pulse received at the
mid-point frequency channel immediately following the mid-point of the
observation.

Our results show that the ability to determine the signal parameters
is unaffected by the choice of data representation when comparing
the Fourier domain approach and the folded data.  This is apparent
from the consistent widths of the posterior distributions which
define the uncertainty in parameter estimation.  The clear effect
that we see is the discrepancy between the location of the posterior
distributions for the case where the error in folding period has been
accounted for and where it has not.  We see that the estimation of the
dispersion, pulse amplitude, pulse width, and pulse frequency is only
marginally affected.  However, the phase parameter and therefore the
time-of-arrival estimate, is strongly biased by the false assumption that
the signal has been folded with the correct pulse period.  For the
results shown in Fig.~\ref{fig:posteriors} the pulse period error of
$10$ nsecs is equivalent to an accumulated phase error of only
$3.6$ degrees over the length of the $25$ sec sub-integrations.
This appears as a $\sim 1.8$ degree ($\equiv 0.005$ cycles) error in
the estimate of $\Phi_{0}$ leading to a $\sim 25$ $\mu$second error in
the estimate of the time-of-arrival value\footnote{This observed phase
  error is half of the total accumulated phase error because the phase
  parameter value is defined at the midpoint of the observation and
  therefore the phase error effectively accumulates over $T/2$ rather
  than $T$.}.

It is clear that the work presented here is intended only as a
potential starting point for more advanced applications of Bayesian
data analysis techniques to the problem of pulsar timing.  A clear
difference between our approach as described here and established
techniques is that we have obtained our pulsar parameter estimates
from a single simulated observation.  The standard approach is to employ a more
global strategy in which the process of producing a time-of-arrival
measurement for a given observation is not just a function of the 
given observation but of all existing observations of the pulsar.  Each TOA represents the reduction of an entire observation
into a single number after having performed a global fit (over all
observations) for a set of
common pulsar parameters e.g. pulse period, the period derivatives,
the sky position, proper motion, pulse shape parameters, the
dispersion measure, etc.  When new observations are taken, the
procedure is repeated and these parameters are refined.  As discussed
in Sec.~\ref{sec:folding}, as data is recorded it is often
reduced (in terms of data volume) by folding at an assumed pulse
period and in addition may be partially de-dispersed with an assumed
dispersion measure.  The detrimental effect of this process (as seen
in our results) will rapidly diminish as more and more observations
are made but further analysis is required to rule out such effects as
contributors to the low-frequency timing noise seen in the msec pulsars.         

The scope of this work is limited to the generation of TOAs but we
would also like to briefly discuss the specific aim of gravitational wave
detection using pulsar timing arrays.  From a purely theoretical
Bayesian data analysis perspective in an ideal scenario, firstly, one would use
an un-reduced dataset spanning all observations of all relevant
pulsars.  Secondly one would construct a model including all pulsar
signal parameters \emph{and} all gravitational wave signal
parameters.  After applying sensible prior distributions to all of
these parameters one would compute marginalised posterior distributions on
both pulsar and gravitational wave parameters and perform model
selection.  We could then establish whether the observations coupled with our prior
beliefs were consistent with the presence of gravitational waves. 
In practice this is a very difficult task for various reasons but most
notably due to the vast computational resources required to process
the vast volume of un-reduced data and to explore the
multi-dimensional parameter space describing the entire pulsar array
and the intervening gravitational wave.  For this reason, in terms of
gravitational wave detection, constructing a reduced dataset is highly
desirable.  In fact, the problem of gravitational wave detection using
timing residuals (the difference between the time-of-arrival values
and those attained by fitting a gravitational wave free pulsar model)
as the initial dataset have already been applied to the specific case
of searching for the gravitational wave stochastic
background~\cite{2009MNRAS.395.1005V,2009PhRvD..79h4030A}.

The apparent separation of the complete gravitational wave detection
problem into a gravitational wave free component, from which a reduced
dataset is produced, and then a second component in which this reduced
dataset is then analysed including the effects of gravitational waves,
seems potentially problematic.  Under the assumption that each TOA
measurement is independent of all others one can argue strongly that
the effect of a low-frequency gravitational wave on each measurement
is negligible and that the TOA truly represents the unambiguous
arrival time of an average pulse within that observation and defined
at some epoch.  As soon as one performs a global fit (neglecting
gravitational waves) over all observations of a given pulsar a
gravitational wave of sufficient amplitude will affect the best fit
pulsar parameters.  Such a procedure could absorb some fraction
of a gravitational wave into the pulsar parameter estimates (e.g the
pulsar period derivatives).  In future work we hope to address this
issue and to provide a comparison between an analysis
using independent TOA measurements as a dataset for gravitational wave
detection and an analysis using globally estimated TOAs.  

In addition we hope to be able to include, and account for, many of the physical
effects and data analysis issues that we have ignored in our toy model
approach.  These include a more robust treatment of the noise where we
allow time and frequency variation and investigate the validity of the
assumption of Gaussianity.  In reference to this we hope to also
include the effects of radio frequency interference (RFI) and
investigate methods in which we are able to analytically marginalise over the noise
and therefore potentially avoid the need to estimate it.   We also aim
to include the effect of polarisation into the analysis.  A
search-mode dataset is itself the product of two independent radio
signal polarisation measurements which are combined as a function of
the Stokes parameters.  These parameters can be incorporated into the
Bayesian framework and uncertainties on these parameters can be
marginalised over in parallel with the signal parameters.  Less well
defined effects to consider include a time and frequency varying
pulse profile parameterisation, time varying dispersion measure,
scattering, scintillation and nulling.  Finally, we hope to develop
this work beyond the toy model to a point at which it can be applied
to real pulsar data.  In such a scenario we will also have to
incorporate barycentric routines~\cite{2006MNRAS.369..655H} to include
the obvious effects of detector motion, sky position uncertainty and,
where applicable, binary orbital motion.  
%
%
\ack

We thank Maura McLaughlin, Benjamin Knispel, Reinhard Prix, Christian
R\"over and Xavier Siemens for insightful discussions and invaluable input.

\section*{References}
\bibliographystyle{unsrt}
\bibliography{masterbib}

\end{document}